\documentclass[11pt]{article}
\usepackage{moriond}

\def\Journal#1#2#3#4{{#1} {\bf #2}, #3 (#4)}

\def\PRL{\em Phys. Rev. Lett.}
\def\PRD{{\em Phys. Rev.} D}

\def\APJ{{\em ApJ}}

\def\MNRAS{{\em MNRAS}}

\def\AA{{\em A\&A}}

\def\be{\begin{equation}}
\def\ee{\end{equation}}
\def\bea{\begin{eqnarray}}
\def\eea{\end{eqnarray}}

\newcommand{\Photo}{\includegraphics[height=35mm]{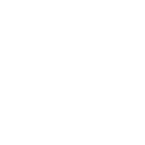}}
\newcommand{\sigmav}{$\langle\sigma v\rangle$ }

\begin{document}
\vspace*{4cm}
\title{Diffuse $\gamma$-ray emission from misaligned active galactic nuclei}
\author{M. Di Mauro$^{\ast}$ $^1$, F. Donato $^1$, F. Calore $^2$}
\address{$^1$ Physics Department, Torino University, and  Istituto Nazionale di Fisica Nucleare, Sezione di Torino, via Giuria 1, 10125 Torino, Italy;
$^2$ Institute for Theoretical Physics, University of Hamburg, Luruper
Chaussee 149, 22761 Hamburg, Germany. 
$^{\ast}$Talk presented by Mattia Di Mauro.}	  

\maketitle

\begin{abstract}
We calculate the diffuse $\gamma$-ray emission due to the population of misaligned AGN (MAGN) unresolved 
by the Large Area Telescope (LAT) on the {\it Fermi} Gamma-ray Space Telescope ({\it Fermi}). 
A correlation between the $\gamma$-ray luminosity and the radio-core luminosity is established and demonstrated to be physical by statistical tests, as well as 
compatible with upper limits based on {\it Fermi}-LAT data for a large sample of radio-loud MAGN. We constrain the derived
$\gamma$-ray luminosity function by means of the source count distribution of the MAGN detected by the {\it Fermi}-LAT. 
We finally estimate the diffuse $\gamma$-ray flux due to the whole MAGN population which ranges from 10\% up to nearly 
the entire measured Isotropic Gamma-Ray Background (IGRB). 
We evaluate also the room left to galactic DM at high latitudes ($>10^\circ$), by taking into account the results on the MAGN together 
with the other significant galactic and extragalactic $\gamma$-rays emitting sources.
\end{abstract}

\section{The correlation between $\gamma$-ray and radio luminosity}
\label{sec:correlation} 

The {\it Fermi}-LAT has measured the Isotropic Gamma-Ray Background (IGRB) with very good accuracy from 200 MeV to 100 GeV  (A. A. Abdo et al. \cite{IDGRB}).
However the nature of the IGRB is still an open problem in astrophysics. 
Blazars, high luminosity Active Galactic Nuclei (AGN) whose jets are oriented along the lines-of-sight (l.o.s.), may contribute to 20\%-30\% of the IGRB (A. A. Abdo et al. \cite{2010ApJ...720..435A}).
AGN with axes misaligned with respect to the line-of-sight (hereafter MAGN) have weaker luminosities but are expected to be more numerous than blazars. It is expected that a non-negligible contribution to the IGRB might be attributable to the unresolved MAGN population.
The aim of this contribution is the estimation of the $\gamma$-ray diffuse emission produced by the cosmological population of MAGN. 
For any details we referer to Di Mauro et al. \cite{dimauro2013}.
\\
The bulk of the $\gamma$-ray radiation from AGN is generated via synchrotron self-Compton (SSC)
or external inverse Compton (EC) scatterings in the central region of the source, the core.
In the absence of predictions for the $\gamma$-ray luminosity function, 
we follow a phenomenological approach to relate the source $\gamma$-ray luminosity to the radio luminosity of the source core. 
The latter is phenomenologically much better established, given the high number of detected MAGN in the radio frequencies. 
\begin{figure*}
\begin{centering}
 \includegraphics[scale=0.6]{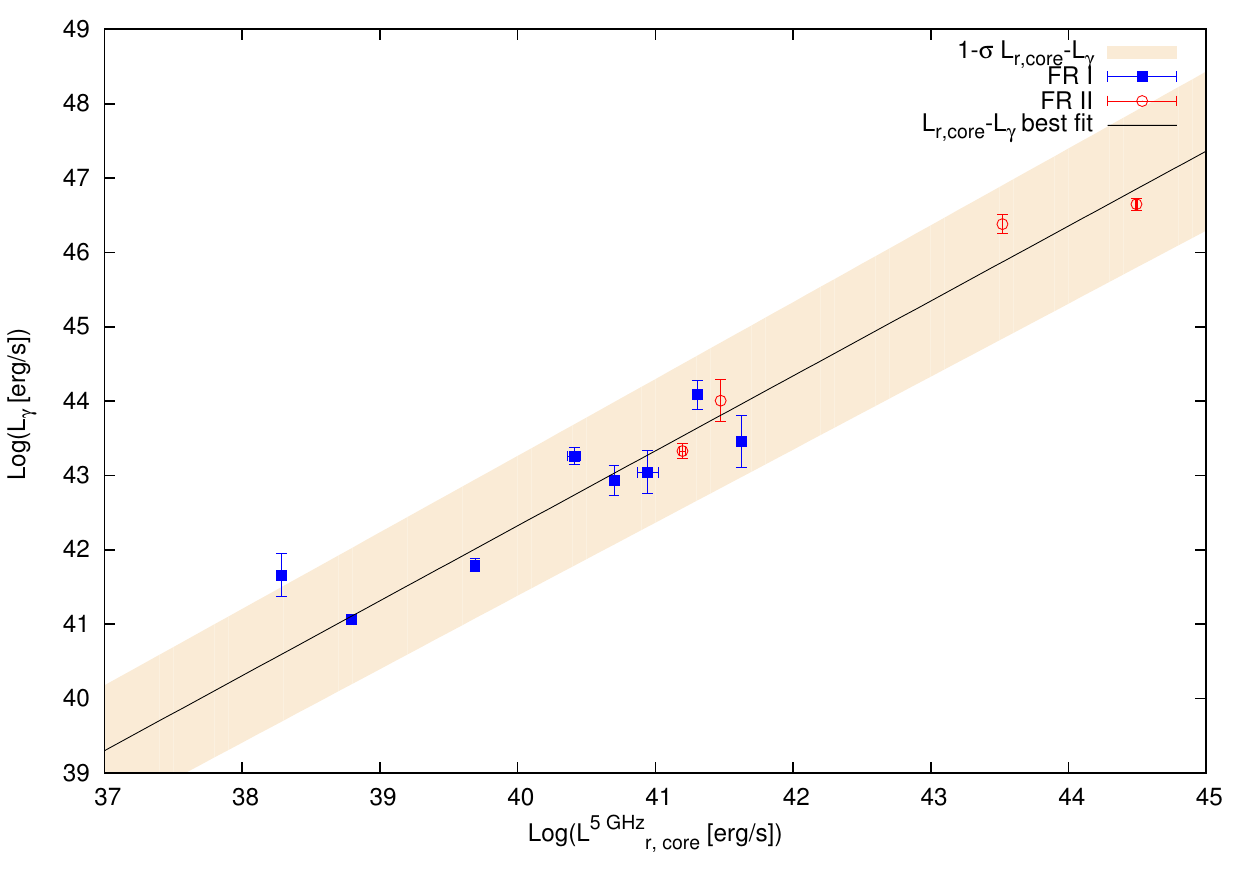}
 \includegraphics[scale=0.6]{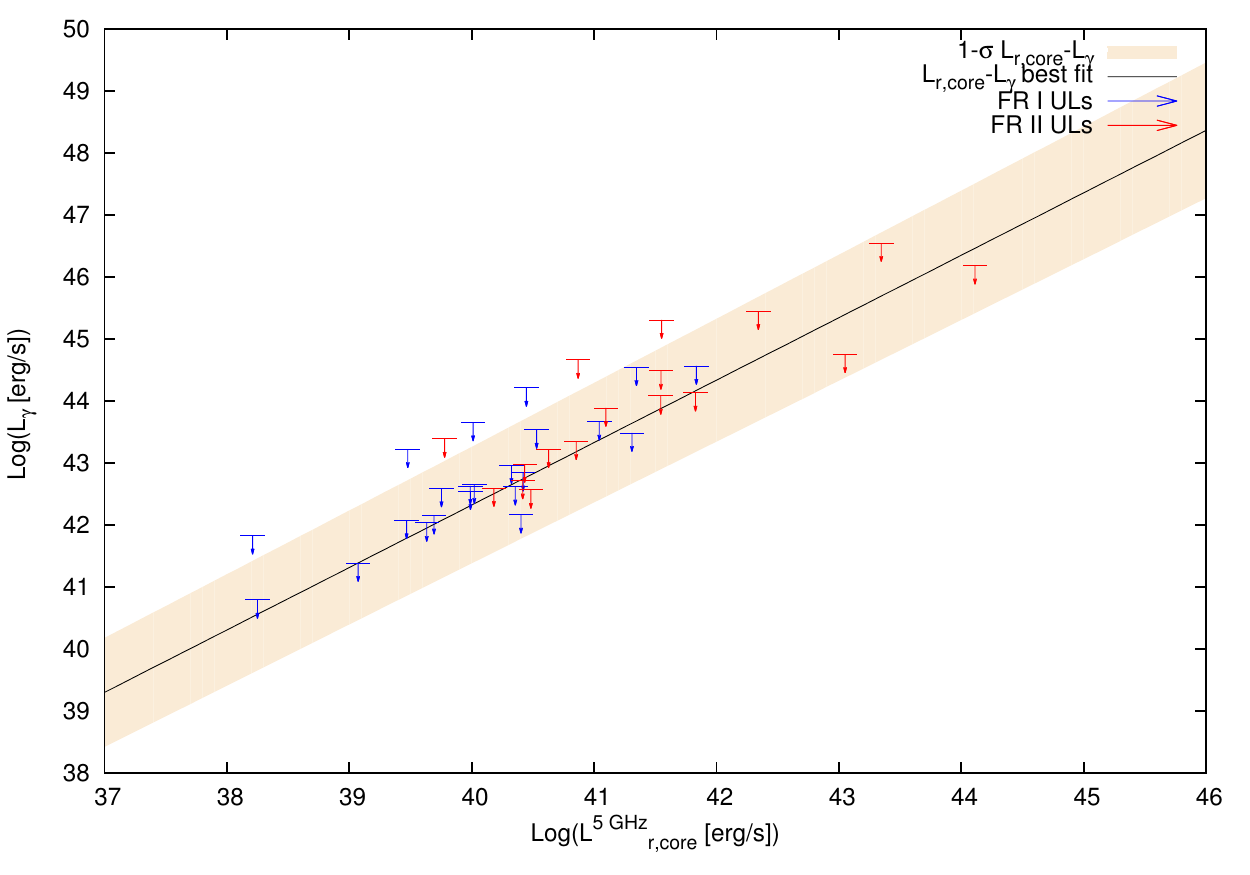}
\caption{On the left panel we display the observed $\gamma$-ray luminosity vs radio core luminosity at 5 GHz for our MAGN sample,
the right panel shows upper limits on a sample of {\it Fermi}-LAT undetected radio-loud MAGN.}
\label{fig:correlation} 
\end{centering}
\end{figure*}
In the first and the second catalogs of LAT AGN sources (A. A. Abdo et al. \cite{misaligned} and M. Ackermann et al. \cite{secondcatalogAGN})
{\it Fermi}-LAT has reported the detection of 15 MAGN.
In Fig.\,\ref{fig:correlation} (left) we plot the core radio and $\gamma$ luminosities for 12 selected MAGN. 
The derived correlation between $L_{r, {\rm core}}$ and $L_\gamma$:
\begin{equation}
     \label{eq:correlation}
         \log{(L_{\gamma})} = 2.00\pm0.98 + (1.008 \pm0.025)\log{ (L^ {5 {\rm GHz}}_{r, {\rm core}} )}
    \end{equation} 
is shown as a solid line and the relevant $1\sigma$ error band as a shaded area.
The correlation could be biased by distance effects and flux-limited samples therefore
we have tested its strength via a Spearman rank-order and a modified Kendall rank correlation test.
We can exclude the correlation happening by chance at the 95$\%$ C.\,L\,.
In order to test the robustness of the core radio $\gamma$-ray luminosity correlation we derive 95$\%$ C.L. $\gamma$-ray upper limits for a sample of 
{\it Fermi}-LAT undetected radio-loud MAGN.
The result in Fig.\,\ref{fig:correlation} indicates that upper limits are consistent with the correlation in Eq.\,\ref{eq:correlation} within its uncertainty band (see Di Mauro et al. \cite{dimauro2013} for details).

\section{The  $\gamma$-ray luminosity function and the source count distribution}
\label{sec:GLF}
The calculation of the diffuse emission from unresolved ($i.e.$ not detected by the {\it Fermi}-LAT) 
MAGN relies on the $\gamma$-ray luminosity function (GLF) for that specific population. 
We derive the GLF from the radio luminosity function (RLF) by exploiting the correlation between radio and $\gamma$-ray luminosities.
We assume that $N_{\gamma}=k\;N_r$, where the normalization $k$ takes into account our ignorance of the number of radio-loud MAGN 
emitting in $\gamma$-rays.
 Therefore the GLF is defined through a RLF by:
 \begin{equation}
     \label{rhogammanew}
        \rho_{\gamma}(L_{\gamma},z) = k \; \rho_{r,{\rm core}}(L^{5 {\rm GHz}}_{r,{\rm core}}(L_{\gamma}),z) 
        \frac{d\log L^{5 {\rm GHz}}_{r,{\rm core}}(L_{\gamma})}{d\log L_{\gamma}}\,,
    \end{equation}
where $\rho_{r,{\rm core}}$ refers to the radio luminosity function of the cores of the MAGN. 
We use the total RLF derived in C. J. Willott et al. \cite{willott} (Model C with $\Omega_M$=0) and obtain the core RLF through the link between 
total and core radio luminosities as in L. Lara et al. \cite{lara2004}.
\\
An important observable for the correctness of our method is provided by the source count distribution of MAGN, i.e.\,the cumulative number of sources 
$N(>F_{\gamma})$ detected above a threshold flux $F_\gamma$ defined as:
    \begin{eqnarray}
    \label{Ncount}
    N_{\rm th}(>F_{\gamma}) &=&  4\pi \; \int^{1.0}_{3.5} \frac{dN}{d\Gamma}  d\Gamma \int^{6}_0 \frac{d^2V}{dz d\Omega} \nonumber
   \int^{10^{50} \rm{erg/s}}_{L_\gamma({F_\gamma},z,\Gamma)}  \; \frac{dL_{\gamma}}{L_{\gamma} \ln(10)} \; \rho_{\gamma}(L_{\gamma},z, \Gamma).
    \end{eqnarray} 
The spectral index distribution, $dN/d\Gamma$, is assumed to be gaussian in analogy with blazars (A. A. Abdo et al. \cite{2010ApJ...720..435A}).
 \begin{figure}
 \begin{centering}
 \label{fig:ncount}
 \includegraphics[scale=0.6]{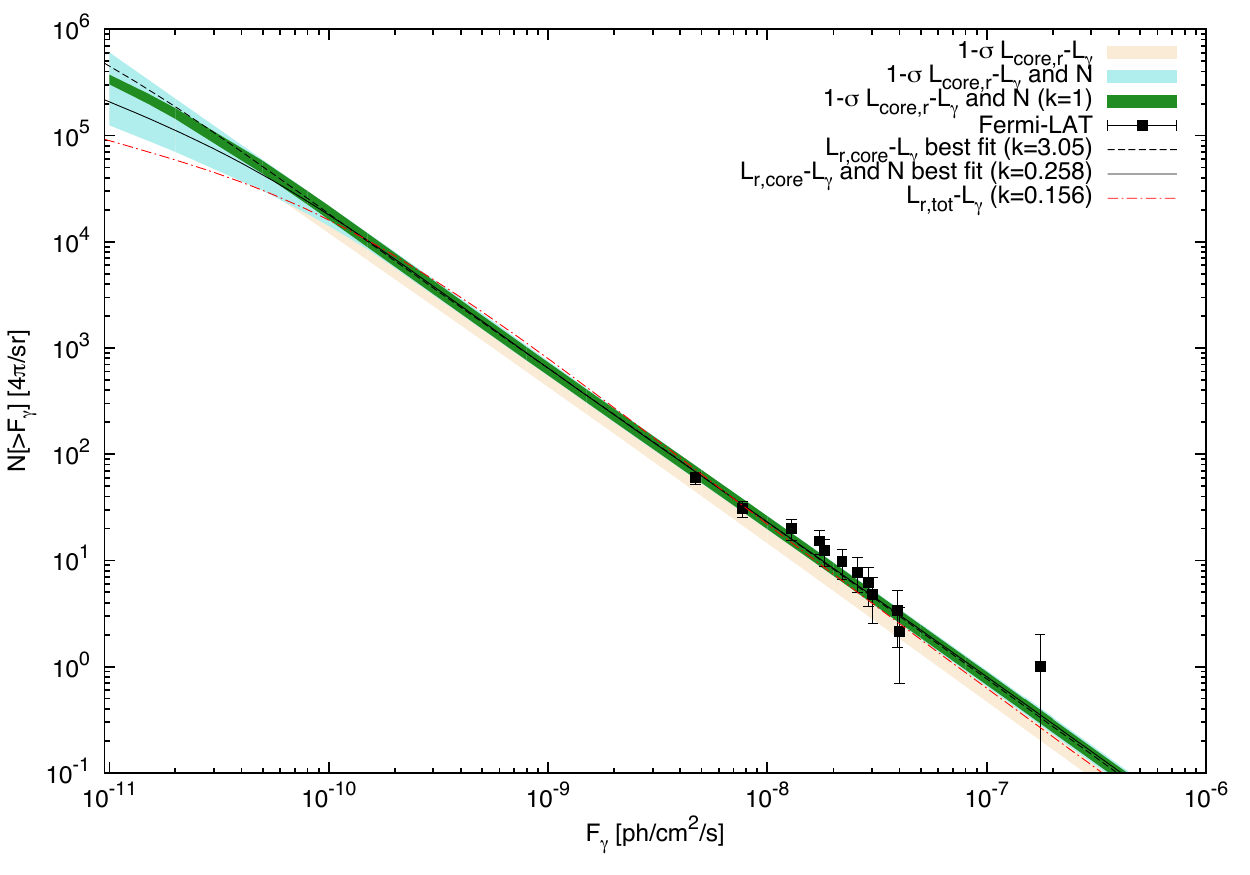}
 \includegraphics[scale=0.6]{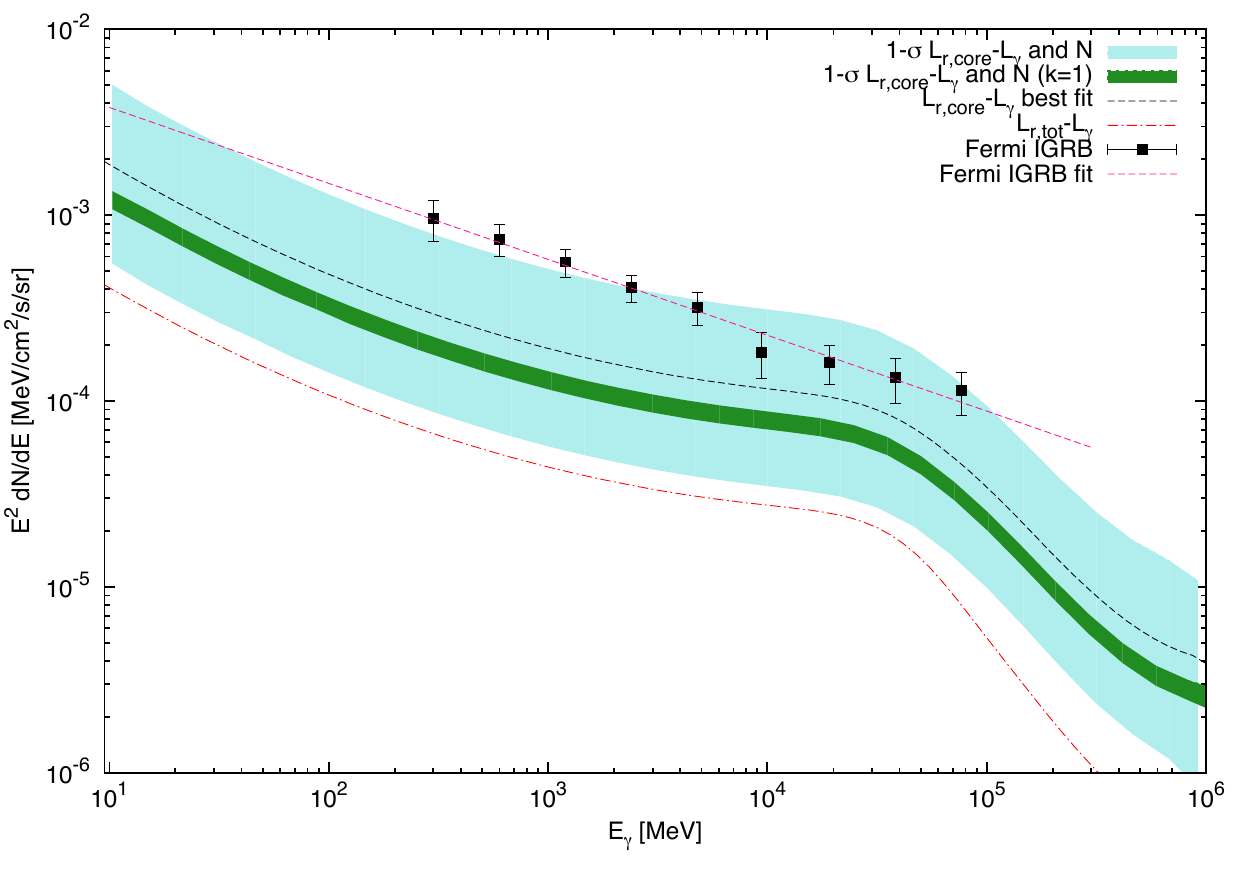}
\caption{On the left panel we show $N_{\rm th}(>F_{\gamma})$ and on the right the diffuse $\gamma$-ray flux due to the MAGN with different bands of uncertainty, overlaid 
 with the experimental data.}
\end{centering}
\end{figure}
Fig.\,\ref{fig:ncount} (left) shows the theoretical $N_{\rm th}(>F_{\gamma})$, with different bands of uncertainty, overlaid 
with the experimental source count distribution (see Di Mauro et al. \cite{dimauro2013} for details).
It is remarkable that the uncertainty bands are in good agreement with the {\it Fermi}-LAT data, supporting the validity of our procedure. 
We stress also that the theoretical source count distribution predicts a large number of MAGN at very low fluxes and this fact may lead to a significant $\gamma$-ray diffuse emission.

\section{The diffuse $\gamma$-ray emission from MAGN}
\label{sec:flux}
The diffuse $\gamma$-ray flux due to the whole population of MAGN may be estimated as follows:
    \begin{eqnarray} 
    \label{egrbdef}
        \frac{d^2F(\epsilon)}{d\epsilon d\Omega} =  \int^{3.5}_{1.0}d\Gamma \frac{dN}{d\Gamma}
	  \int^{6}_0 \frac{d^2V}{dz d\Omega} \int^{10^{50} \rm{erg/s}}_{10^{41} \rm{erg/s}} \frac{d F_{\gamma}}{d\epsilon} 
	   \frac{dL_{\gamma}}{L_{\gamma} \ln(10)}\rho_{\gamma}(L_{\gamma},z) (1-\omega({F_{\gamma}}))\exp{(-\tau_{\gamma,\gamma}(\epsilon,z))}.  \nonumber
    \end{eqnarray}
The term $\omega(F_{\gamma,i})$ is the detection efficiency of the {\it Fermi}-LAT, 
$d F_{\gamma}/d\epsilon$ is the intrinsic photon flux at energy $\epsilon$ and $\tau_{\gamma,\gamma}$ is the optical depth of $\gamma$ rays
($\epsilon > 20$ GeV) propagating in the Universe and absorbed by the interaction with the extragalactic background light (EBL).
Fig.\,\ref{fig:ncount} (right) shows the diffuse $\gamma$-ray flux due to the MAGN population with the relevant uncertainty band, 
which is nearly a factor of ten wide, as a function of $\gamma$-ray energy and along with the {\it Fermi}-LAT data for the IGRB (A. A. Abdo et al. \cite{IDGRB}).
The effect of EBL absorption is clear from the softening of the flux above 50 GeV.
At all {\it Fermi}-LAT energies, the best fit MAGN contribution is 20\%-30\% of the measured IGRB flux.
The intensity from MAGN integrated above 100 MeV is $9.83\cdot 10^{-7}/2.61\cdot 10^{-6}/8.56\cdot 10^{-6}$ photons cm$^{-2}$ s$^{-1}$  sr$^{-1}$, 
when considering the lower/best fit/upper curve of the uncertainty band. 
These numbers represent 9.5\%/25\%/83\% of the IGRB, respectively. 
The analogous calculation for the two blazar populations of BL Lacs and 
FSRQs and for star-forming galaxies (SF) gives respectively $7.83^{+1.09}_{-2.34}\cdot 10^{-7}$ photons cm$^{-2}$ s$^{-1}$  sr$^{-1}$ ($\sim$ 8\% of the IGRB), 
$9.66^{+1.67}_{-1.09} 10^{-7}$ photons cm$^{-2}$ s$^{-1}$  sr$^{-1}$ ($\sim$ 10\%) and 
$8.19^{+7.31}_{-3.89} \cdot 10^{-7}$ photons cm$^{-2}$ s$^{-1}$ sr$^{-1}$ ($\sim$ 8\%) (see F. Calore et al. \cite{2012PhRvD..85b3004C} and refs. therein).

\section{Constraints on Dark Matter annihilation into $\gamma$-rays}
\label{sect:constraints}
The self-annihilation of DM pairs in the haloes of galaxies may give birth, among other species, to
$\gamma$-rays. 
\begin{figure*}
\begin{center}
\label{fig:upperlimits}
\includegraphics[scale=0.60]{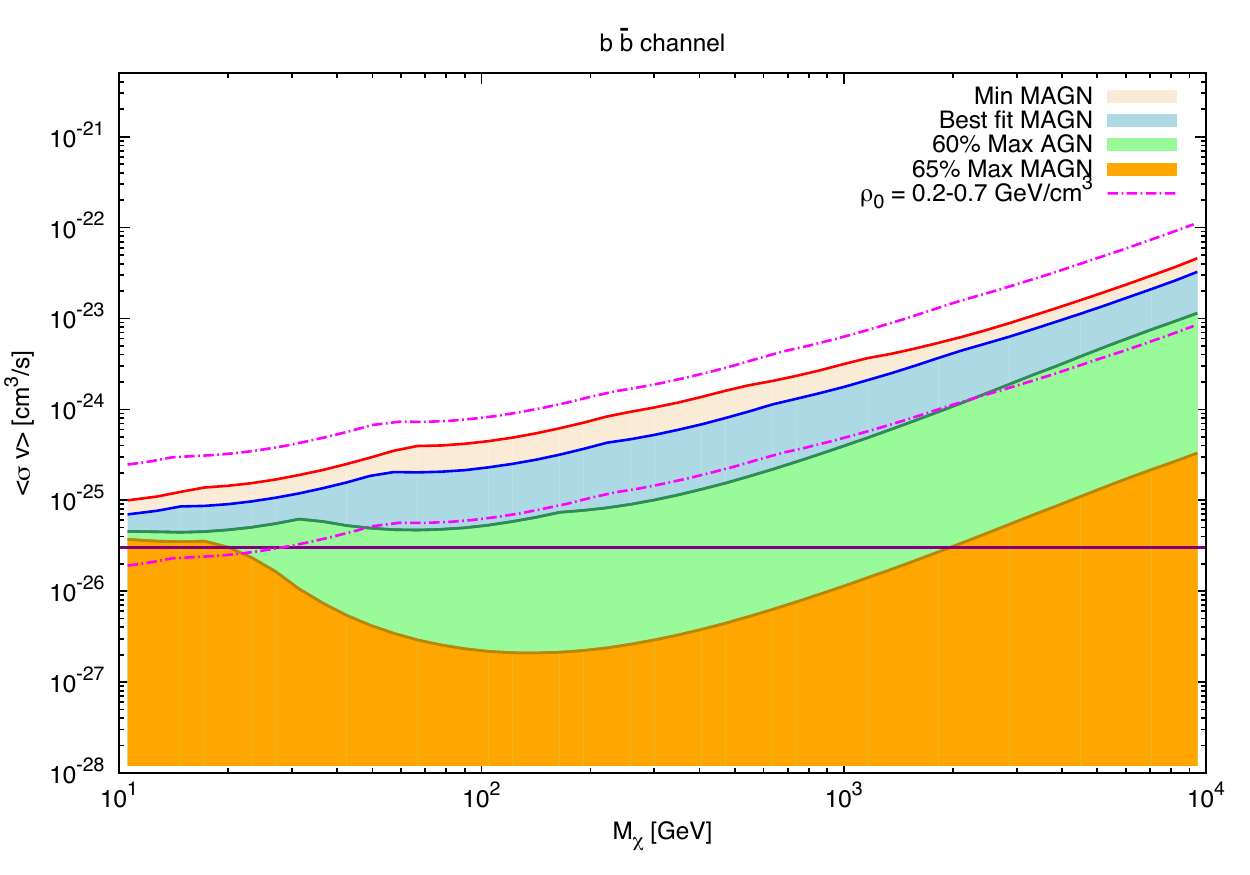}
\includegraphics[scale=0.60]{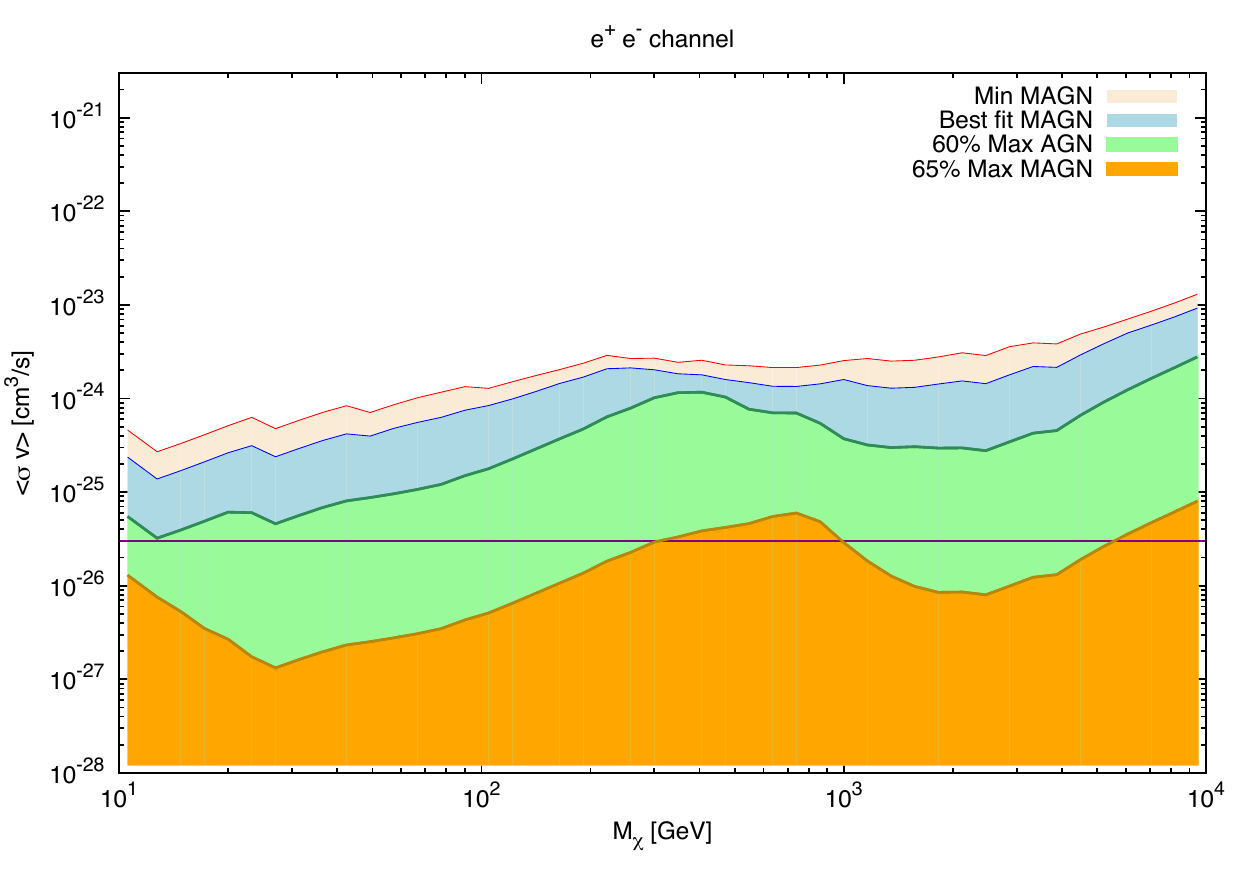} 
\caption{Upper limits to \sigmav into $\gamma$-rays from
$b\bar{b}$ (left) and $e^+e^-$(right). The MAGN flux is varied from the minimum predicted value (red upper curve),  to its best fit (cyan), to 60\% (green) 
and 65\% (orange, bottom curve) of the predicted maximum level.  Pink dashed lines illustrate the uncertainty on the DM local density.}
\end{center}
\end{figure*}
We have calculated the direct (prompt emission) and indirect production (through Inverse Compton scattering), of
 $\gamma$-rays produced by WIMP pair annihilation in the halo of our Galaxy. 
We considered a Burkert DM profile for latitudes $|b|>10^\circ$ with a local DM density of
$\rho(r=r_{\odot})=0.4$ GeV cm$^{-3}$ and $r_\odot$=8.33 kpc.
In Fig.\,\ref{fig:upperlimits} we display the upper bounds on the DM annihilation cross section averaged on the velocity distribution \sigmav for annihilation 
channels into $b\bar{b}$ (left) and $e^+e^-$ (right).
In order to appreciate how the uncertainty on the MAGN flux prediction affects the bounds on \sigmav, we have fixed the diffuse 
contribution from BL Lacs, FSRQs, SF galaxies and millisecond pulsar (MSPs)
at their best fit values and varied the level of the MAGN flux from its minimal predicted value up to 60\% and 65\% of the maximum flux estimated. 
Were the MAGN contribute about 60\%-65\% of the predicted maximal flux,
the constraints on \sigmav would be set  below the thermal decoupling cross section $2 \cdot 10^{-26}$ cm$^3$/s expected for a generic WIMP DM candidate,
depending on the mass of the DM particle and on the annihilation channels (see Calore et al. \cite{calore2013} for details).

\section{Conclusions}
\label{sec:conclusions}
We have calculated the $\gamma$-ray flux from the MAGN cosmological population. We have first established the existence (at 95\,\% C.L.) of a correlation between the radio core and the $\gamma$-ray luminosities of the MAGN detected by the {\it Fermi}-LAT. 
We then used this correlation to infer the $\gamma$-ray luminosity function from a well established radio luminosity function, and further tested the former against the MAGN count distribution measured by the {\it Fermi}-LAT. 
Using our $\gamma$-ray luminosity function, and after taking into account $\gamma$-ray absorption from EBL, we have predicted the diffuse $\gamma$-ray flux due to MAGN and estimated an uncertainty band of about a factor ten. The best fit to the MAGN diffuse $\gamma$-ray emission is 20\%-30\% of the measured IGRB flux.
Moreover we have found that the cosmological population of faint and numerous MAGN, 
when added to the contribution from other sources ($i.e.$ blazars, SF galaxies, MSPs), could entirely explain the observed IGRB. 
We have demonstrated that this scenario would leave very little room to more exotic sources, such as DM in the halo of our Galaxy.

\end{document}